\documentclass[a4paper]{jpconf}
\usepackage{graphicx}
\begin{document}
\title{Top quark and Electroweak measurements at the Tevatron}

\author{Lucio Cerrito}

\address{{\it Department of Physics, Queen Mary University of London, London, UK} \\ on behalf of the CDF and D0 Collaborations }

\ead{l.cerrito@qmul.ac.uk}
\begin{abstract}
We present recent preliminary measurements at the Tevatron of $t\bar{t}$ and single top production cross section, top quark mass and width, top pair spin correlations and forward-backward asymmetry. In the electroweak sector, we present the Tevatron average of the $W$ boson width, and preliminary measurements of the $W$ and $Z$ forward-backward asymmetries and $WZ,~ZZ$ diboson production cross sections. All measurements are based on larger amount of collision data than previously used and are in agreement with the standard model. 
\end{abstract}

\section{Introduction}
The Tevatron accelerator complex produces $p\bar{p}$ collisions with a center of mass energy of $\sqrt{s}=1.96~{\rm TeV}$, and the CDF and D0 experiments have so far recorded $\sim8$~fb$^{-1}$ of collision data. Just a fraction of this data allowed establishing a successful experimental programme of physics of the top quark and of the electroweak sector. In the following, we summarise the main measurements and report on preliminary results based on the latest data analyses.

\section{Top quark pair and single top production}
\begin{figure}[h]
\begin{minipage}{18pc}
\includegraphics[width=18pc]{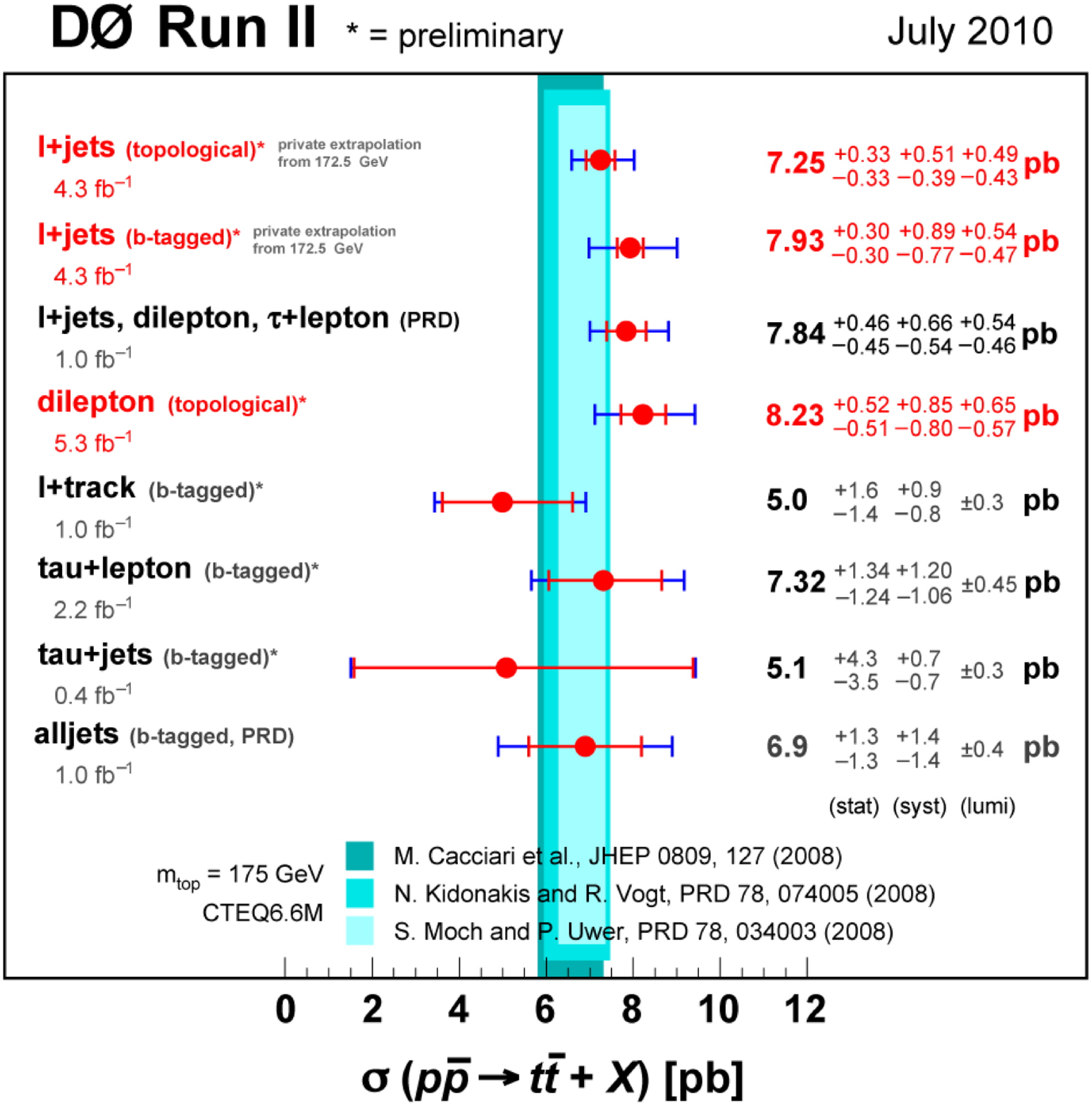}
\caption{\label{fig:d0_top_pair_xsec}Summary of $t\bar{t}$ production cross section measurements using samples in different decay channels recorded with the D0 detector.}
\end{minipage}\hspace{2pc}
\begin{minipage}{18pc}
\includegraphics[width=18pc]{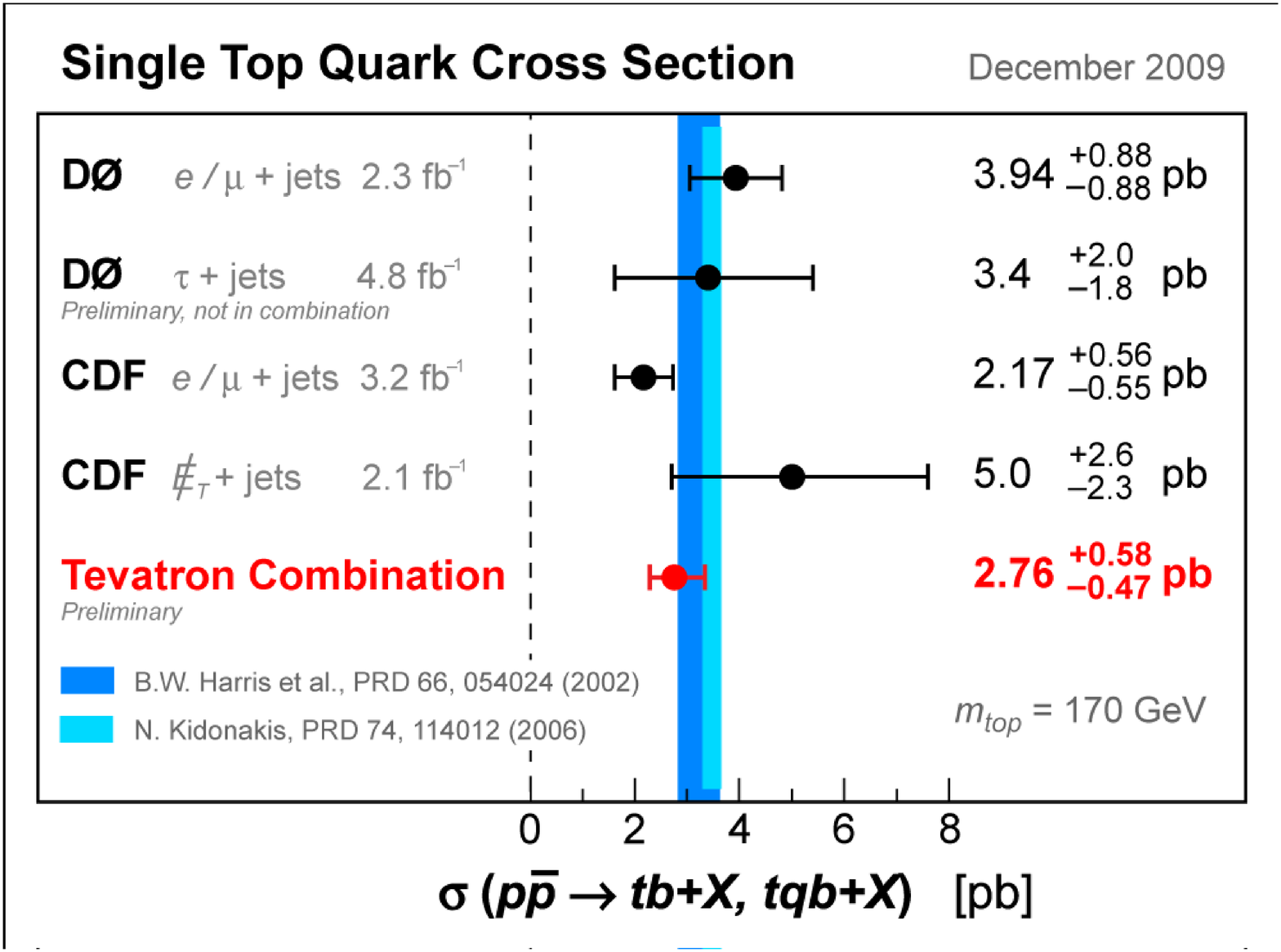}
\caption{\label{fig:singletop}Measurements of single top cross section performed with data collected by the CDF and D0 collaborations. Also indicated the preliminary Tevatron combination.}
\end{minipage}
\end{figure}

Top quarks are produced and observed at the Tevatron in their pair and singlet modes. Measurements of the cross section allow testing of the production and decay mechanisms of the top and its coupling in the standard model (SM). The pair production of top and anti-top in $p\bar{p}$ collisions has been measured with increasing precision since its discovery in 1995 \cite{top_discovery,top_discovery2} at the Tevatron. Figure \ref{fig:d0_top_pair_xsec} shows the most recent measurements using data collected by the D0 experiment \cite{d0detector} in various decay channels. As top quarks decay almost exclusively to $W+b$, the final states of $t\bar{t}$ events are labelled as {\it dilepton}, {\it lepton plus jets} and {\it all-hadronic} depending on whether a leptonic decay has occurred in both, one only or none of the two final-state $W$ bosons respectively. All results are in agreement with each other and in agreement with the next-to-leading order (NLO) calculation for $t\bar{t}$ production in $p\bar{p}$ collisions at $\sqrt{s}=1.96$~TeV of 7.5$\pm$0.8~pb \cite{theoxsec1,theoxsec2,theoxsec3}. Preliminary measurements are also shown in figure \ref{fig:d0_top_pair_xsec}, performed with 4.3--5.3~fb$^{-1}$ of collision data, and yield individually a precision of $\sim 11$\%, similar to the theoretical uncertainty. Since a large component (about 6\%) of the  experimental uncertainty on the $t\bar{t}$ cross section is due to the uncertainty on the luminosity determination, higher precision can be obtained by measuring the ratio of $t\bar{t}$ to $Z$-boson cross sections $\sigma_{tt}/\sigma_{Z}$ \cite{cdf_ttbarxsec} and using the theoretical $Z$ boson cross section calculation as input. With such an approach, the uncertainty on the luminosity largely cancels out and the most accurate preliminary determination of the $t\bar{t}$ cross section, combining samples in all decay channels seen with the CDF~II detector \cite{cdfdetector1,cdfdetector2} yields: $\sigma_{t\bar{t}}=7.50\pm0.31({\rm stat.})\pm0.34({\rm syst.})\pm0.15({\rm lumi.})$~pb (for a top mass of $m_t=172.5~GeV/c^2$), in excellent agreement with NLO calculations and with an experimental uncertainty of only $\pm6.4$\%.

The electroweak production of single top quarks has been established at the Tevatron in 2009 \cite{singletop_cdf,singletop_d0}. A preliminary measurement using 4.8~fb$^{-1}$ of collision data recorded with the D0 detector has now been completed in the $\tau$+jets decay channel, measuring $\sigma_{tbX+tqbX}=3.4^{+2.0}_{-1.8}$~pb. Figure \ref{fig:singletop} shows the current best measurements of single top cross section ($s$-channel and $t$-channel combined) and the preliminary Tevatron combined cross section of $2.76^{+0.58}_{-0.47}$~pb, in agreement with the NLO theoretical prediction of $3.4\pm0.2$~pb \cite{stopth1,stopth2,stopth3} (for $m_t=170~GeV/c^2$). This also allows extraction of the Cabibbo-Kobayashi-Maskawa matrix element $|V_{tb}|=0.88\pm0.07$ with a 95\% confidence level (C.L.) lower limit of $|V_{tb}|>0.77$.

\section{Top quark mass and properties}
\begin{figure}[h]
\begin{minipage}{20pc}
\includegraphics[width=12pc]{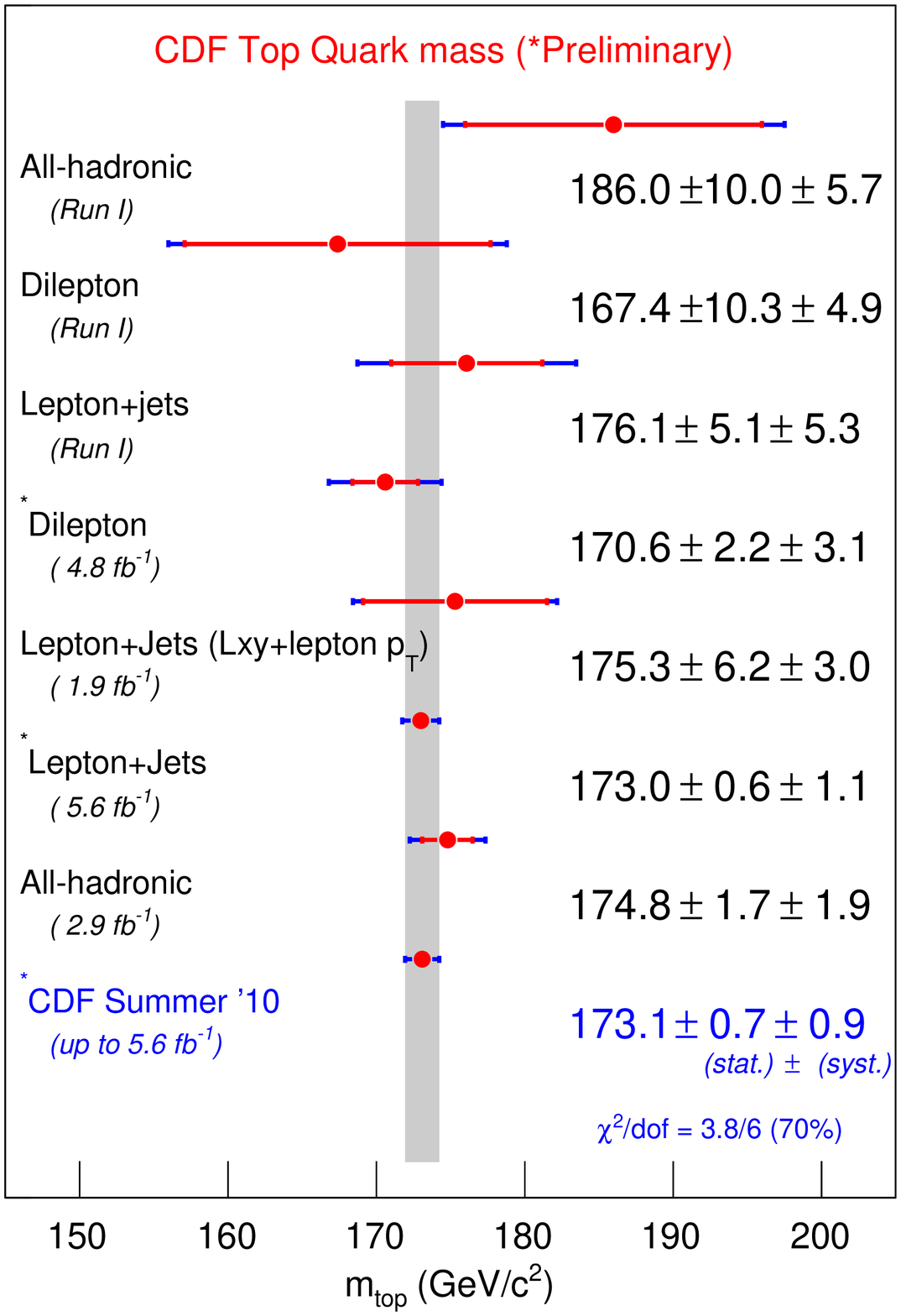}
\caption{\label{fig:topmass}Preliminary and published best measurements of the mass of the top quark in each decay channel, based on data from Run~I (1992-1996) and Run~II (2001-present) analysed by the CDF collaboration. Shown also is the average of those measurements.}
\end{minipage}\hspace{2pc}
\begin{minipage}{16pc}
\includegraphics[width=17pc]{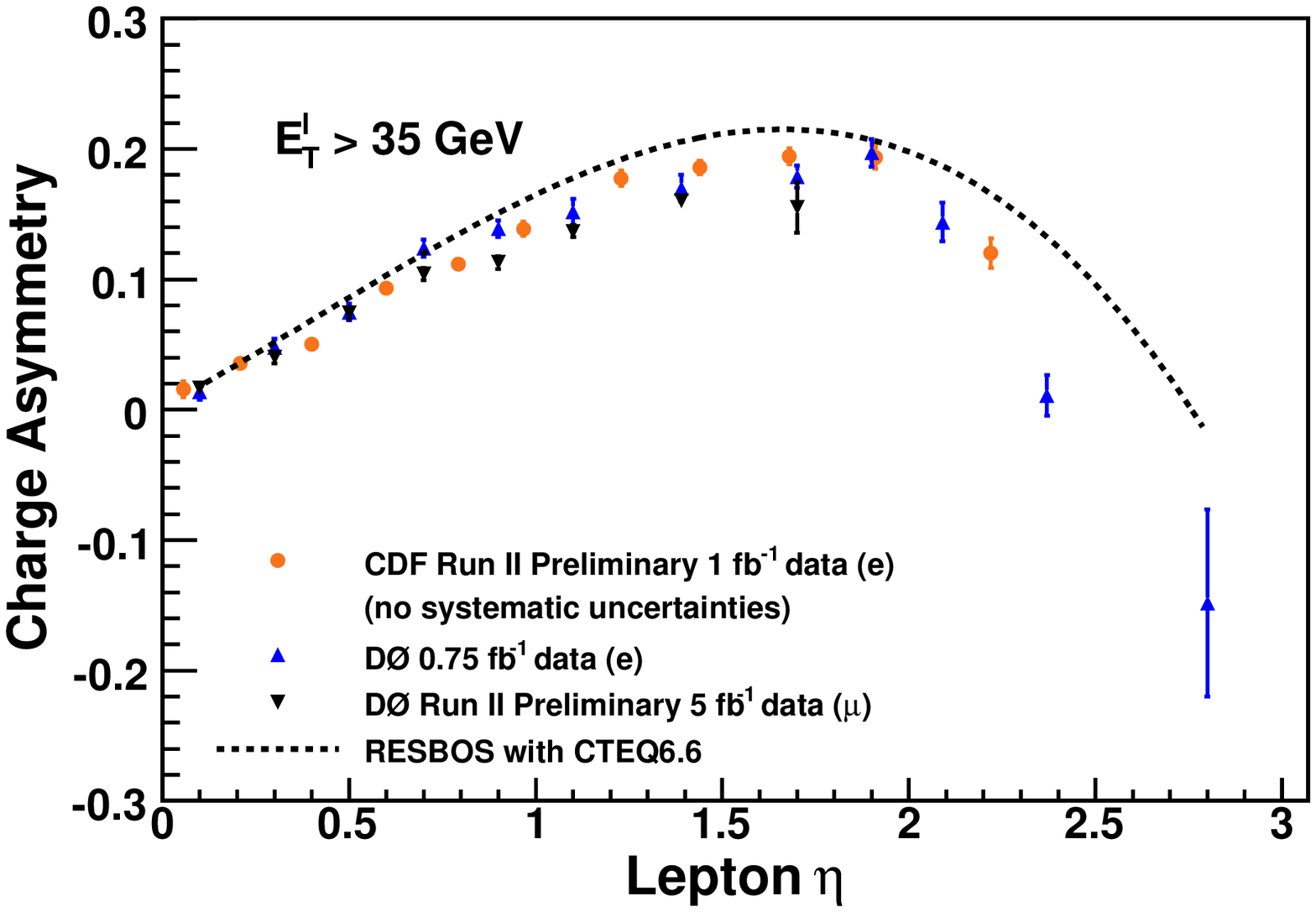}
\caption{\label{fig:wasymm}Comparison of the $W$ boson charge asymmetry measured by the CDF and D0 collaborations as a function of lepton $|\eta|$, for lepton's transverse energies $E_T^\ell>35~{\rm GeV}$. The data are compared to a simulation based on {\tt RESBOS} with {\tt CTEQ6.6} PDFs.}
\end{minipage}\hspace{2pc}
\end{figure}

The mass of the top quark is of interest both because the top is the heaviest known fundamental particle and because a precise measurement helps constraining the mass of the Higgs boson. We present two preliminary measurements using events in the dilepton and lepton plus jets final states reconstructed with the CDF~II detector. The analyses were based respectively on 4.8 fb$^{-1}$ and 5.6 fb$^{-1}$ of collision data and yield experimental top mass values of $m_t=170.6\pm2.2({\rm stat.})\pm3.1({\rm syst.})$~GeV/$c^2$ and $m_t=173.0\pm0.6({\rm stat.})\pm1.1({\rm syst.})$~GeV/$c^2$. Figure \ref{fig:topmass} shows the CDF combination of top mass measurements, $m_t=173.1\pm0.7({\rm stat.})\pm0.9({\rm syst.})$~GeV/$c^2$, which with an accuracy of $0.67\%$ is more precise than the previous world average \cite{topmasscombo09}. Moreover, measurements derived using all the $t\bar{t}$ decay channels are consistent with each other, and a preliminary determination of the top and anti-top mass difference using 5.6~fb$^{-1}$ of collision data recorded with the CDF~II detector gives $m_t-m_{\bar{t}}=-3.3\pm1.4({\rm stat.})\pm1.0({\rm syst.})$~GeV/$c^2$. Taking into account the preliminary results quoted above, the new Tevatron combined experimental top mass determination is $m_t=173.3\pm1.1$~GeV/$c^2$ \cite{topmasscombo10}. Projections indicate that the Tevatron experiments should reach individually a precision of $\pm$0.8--1.0~GeV/$c^2$ with $\sim10$~fb$^{-1}$ of data.

The experimental study of physics of top quarks at the Tevatron includes the determination of a variety of properties in addition to the production cross section and the top quark mass. Measurement of properties such as the production mechanism, the forward-backward asymmetry and differential cross sections for example are only recently approaching the accuracy needed to test the SM predictions. We present two preliminary results on the width of the top quark. In the first, using 4.3~fb$^{-1}$ of Tevatron's $p\bar{p}$ collision data collected by the CDF~II detector, we exploit the reconstructed top quark mass to set an upper limit on the top quark width ($\Gamma_{top}$) of $\Gamma_{top}<7.6~{\rm GeV}/c^2$ at 95\% C.L. We also set a central limit of $0.3~{\rm GeV}/c^2<\Gamma_{top}<4.4~{\rm GeV}/c^2$ at 68\% C.L. In the second result, we extract the total width of the top quark from the partial decay width $\Gamma(t\rightarrow Wb)$ and the branching fraction $BR(t\rightarrow Wb)$. $\Gamma(t\rightarrow Wb)$ is obtained from the measured $t$-channel cross section for single top quark production in 2.3 fb$^{-1}$ of $p\bar{p}$ data from the D0 collaboration, and $BR(t\rightarrow Wb)$ is extracted from a measurement of the ratio $R=BR(t\rightarrow Wb)/BR(t\rightarrow Wq)$ in $t\bar{t}$ events in 1 fb$^{-1}$ of integrated luminosity. Assuming $BR(t\rightarrow Wq)=1$, where $q$ includes any kinematically accessible quark, the result is: $\Gamma_t=2.1\pm0.6~{\rm GeV}/c^2$ which corresponds to a top quark lifetime of $\tau_t=(3\pm1)\cdot 10^{-25}$~s. 

A preliminary measurement of the top quark charge, based on 2.7~fb$^{-1}$ of collision data collected with the CDF~II detector and using soft electron and muon $b$-tagging, now excludes a top quark charge of $-4e/3$ with 95\% C.L, extending the significance of a previous result. Measurements of the dynamics of production and decay of the top quark are of importance because they are sensitive to new production and decay mechanisms. For example, SM top pair production induces a characteristic spin correlation which can be modified by new production mechanisms such as $Z^\prime$ bosons or Kaluza-Klein gluons \cite{kkgluon}. We report on a preliminary measurement of $t\bar{t}$ helicity fractions and spin correlation in 5.3~fb$^{-1}$ of reconstructed lepton+jets data with the CDF~II detector. Choosing the direction of the top quark momentum as the spin quantization axis (the {\it helicity basis}), we find a spin correlation coefficient $\kappa_{hel}=0.48\pm0.48({\rm stat.})\pm0.22({\rm syst.})$, in agreement with the NLO expectation of $\kappa\simeq 0.35$ \cite{spincorrpred}. We also report a preliminary measurement of the forward-backward asymmetry of pair produced top quarks in 5.3~fb$^{-1}$ of data collected with the CDF~II detector. A small asymmetry of 0.038$\pm$0.006 in the laboratory rest frame is expected in Quantum Chromo Dynamics at NLO \cite{afb} and large positive asymmetries with large uncertainties were measured at CDF and D0 \cite{afbcdf,afbd0}. We find $A_{lab}=0.15\pm0.05({\rm stat.})\pm0.02({\rm syst.})$, in agreement with the SM.

\section{\boldmath $W$ boson mass and width}
A precise measurement of the $W$ boson mass ($m_W$) provides a measurement of quantum loop corrections, and in the SM these corrections to $m_W$ are contributed predominantly by the top quark and the Higgs boson loops. Therefore, $m_W$ in conjunction with the top quark mass constrains the mass $m_H$ of the undiscovered Higgs boson. The most recent combination of $m_W$ measurements at the Tevatron is reported in detail elsewhere \cite{wmasstev}, and when combined with measurements at the LEP collider yields the value $m_W=80,399\pm23~{\rm MeV}/c^2$ \cite{lepewwg}. Overall, precision electroweak measurements, including the preliminary value presented in this article of $m_t=173.3\pm1.1~{\rm GeV}/c^2$, indicate the preferred Higgs boson mass of $m_H=89^{+35}_{-26}~{\rm GeV}/c^2$, or $m_H<185~{\rm GeV}/c^2$ at 95\% C.L.

In addition to the $W$ boson mass, the decay width of the $W$ is also of importance towards a stringent test of the SM and can be used to constrain other SM parameters, such as the $V_{cs}$ Cabibbo-Kobayashi-Maskawa matrix element. A recent combination of the CDF and D0 Run~I and Run~II measurements, which are based on fits to tails of the $W$ transverse mass \cite{wmt} distributions, yields $\Gamma_W=2,046\pm49~{\rm MeV}/c^2$ \cite{wwidthcombo}, and a preliminary world average including both the Tevatron and LEP2 gives $\Gamma_W=2,085\pm42~{\rm MeV}/c^2$.

\section{\boldmath $W$ and $Z$ charge asymmetries}
The production of $W$ bosons at the Tevatron is subject to a well known charge asymmetry, due to the different fraction of the proton momentum carried on average by $u$ and $d$ quarks:
\begin{equation}
A(y_W)=\frac{d\sigma^+/dy_W-d\sigma^-/dy_W}{d\sigma^+/dy_W+d\sigma^-/dy_W}
\end{equation}
where $y_W$ is the $W$ boson rapidity \cite{rapidity} and $d\sigma^\pm/dy_W$ is the differential cross section for $W^+$ or $W^-$ boson production. Previous measurements \cite{wasyD0, wasyCDF} have been published by the D0 and CDF collaborations, based on candidate $W\rightarrow e\nu$ events in 0.75~fb$^{-1}$ and 1~fb$^{-1}$ of collision data respectively, in which comparisons to Monte Carlo (MC) simulations of $W$ production and decay showed generally good agreement. We present a direct comparison of these analyses with an additional preliminary measurement based on candidate $W\rightarrow \mu\nu$ events in 4.9~fb$^{-1}$ of collision data collected with the D0 detector. The data are compared to MC event simulation from {\tt RESBOS} \cite{resbos} with {\tt CTEQ6.6} \cite{cteq} parton distribution functions (PDFs), as a function of the lepton pseudo-rapidity $\eta$, in different ranges of the lepton's transverse energy ($E_T^\ell$). While in the low $E_T^\ell$ the data are well reproduced by the MC, for $E_T^\ell>35~{\rm GeV}$ discrepancies start to emerge and the MC is now seen to overestimate the asymmetry at large $|\eta|$ (see figure \ref{fig:wasymm}) with respect to both CDF and D0 experimental data. 

\begin{figure}[h]
\begin{minipage}{16pc}
\includegraphics[width=16pc]{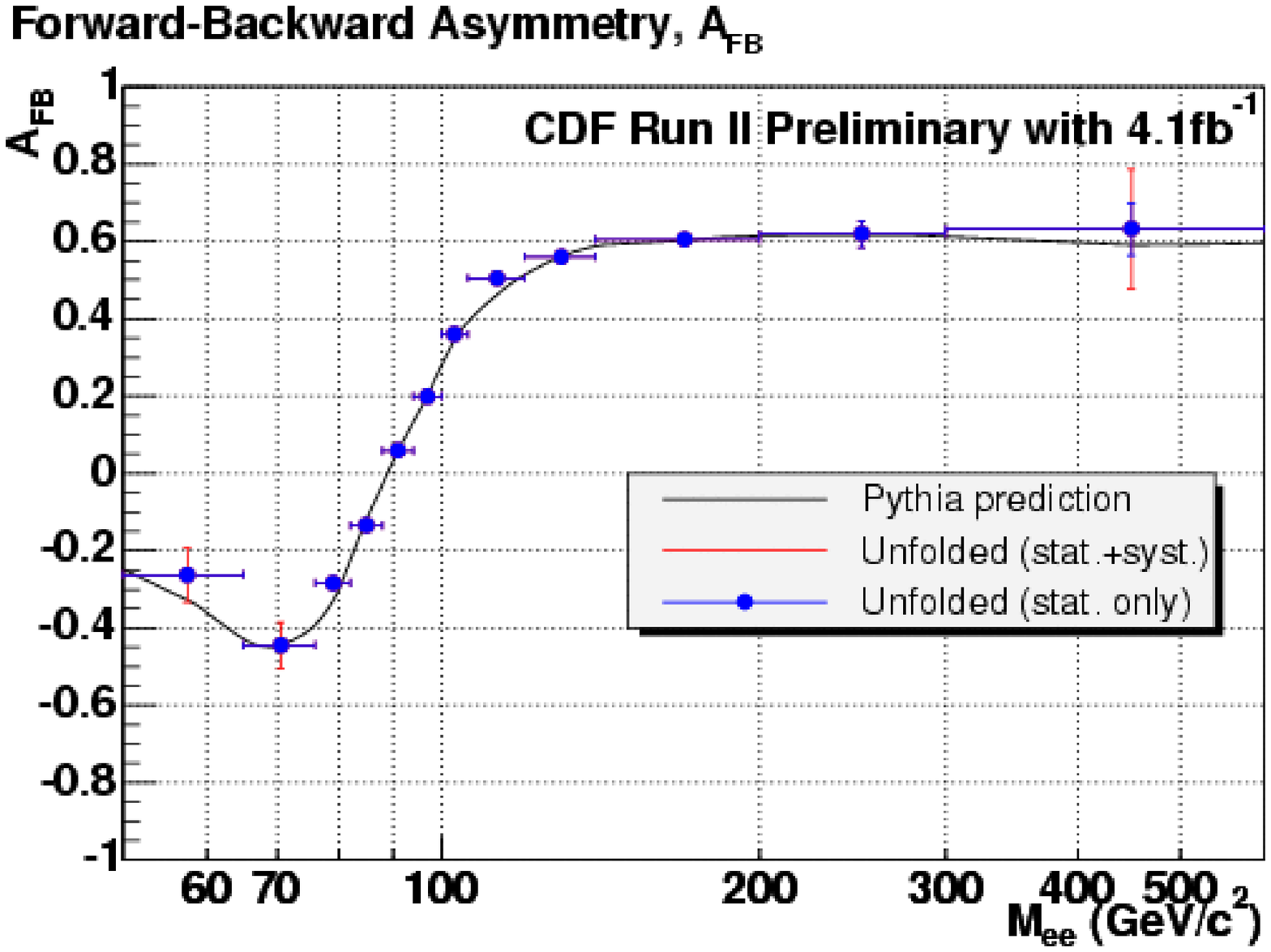}
\caption{\label{fig:zfb} Forward-backward charge asymmetry of high mass Drell-Yan dielectron pairs.}
\end{minipage}\hspace{2pc}
\begin{minipage}{20pc}
\includegraphics[width=15pc]{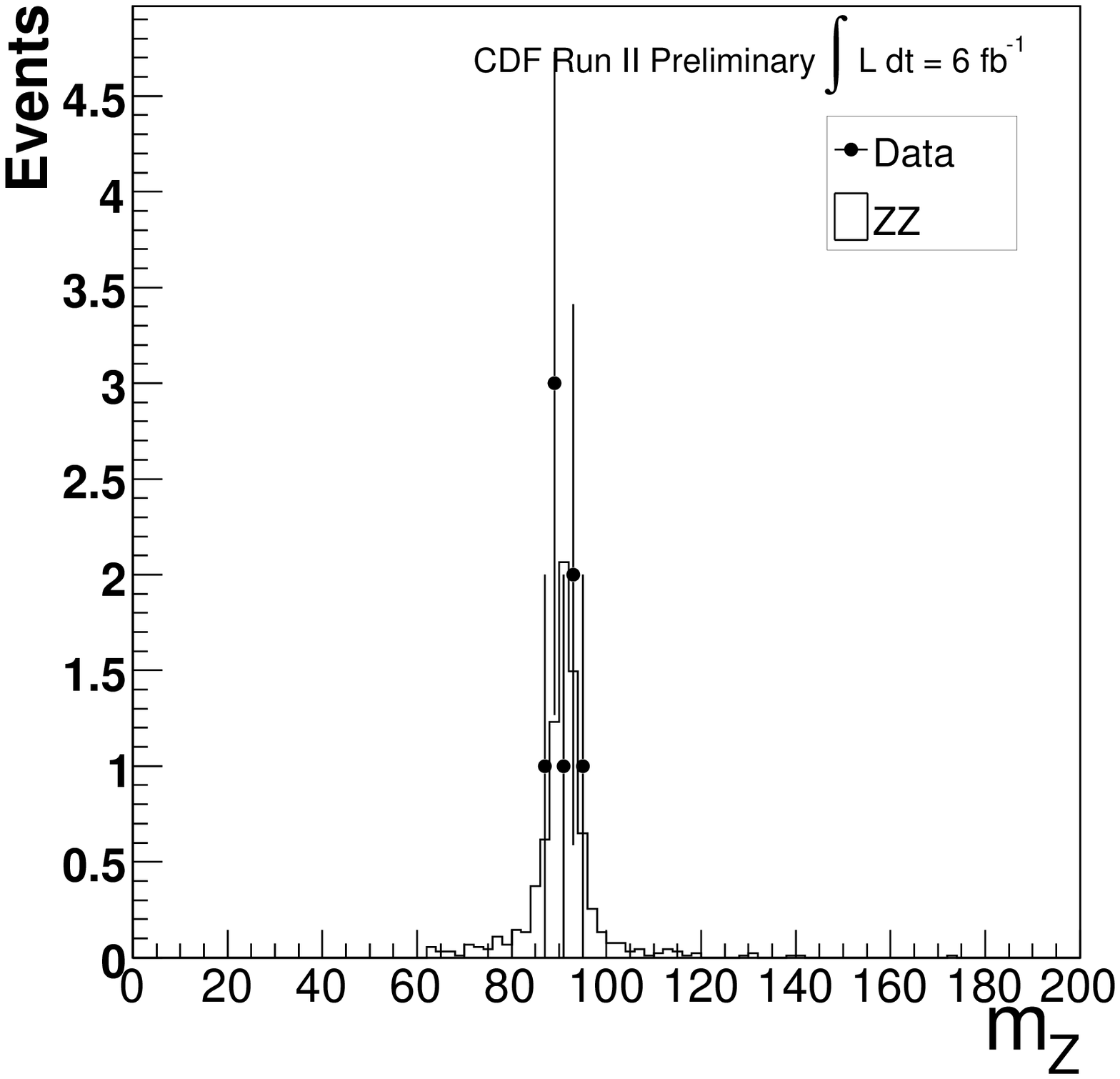}
\caption{\label{fig:zzmass} Invariant mass for the $Z$ candidates reconstructed in the 4 $ZZ$ candidate events.}
\end{minipage}\hspace{2pc}
\end{figure}

The presence of both vector and axial-vector couplings in $q\bar{q}\rightarrow Z/\gamma^*\rightarrow\ell^+\ell^-$ gives rise to an asymmetry also for $Z$ bosons in the polar angle of the negatively charged lepton momentum relative to the incoming quark direction in the rest frame of the lepton pair. We present a preliminary measurement of such forward-backward asymmetry in the kinematic region where the invariant mass ($M_{ee}$) of the lepton pair is between 50 and 600~GeV/$c^2$. The data sample consists of 4.1~fb$^{-1}$ of $p\bar{p}$ collisions recorded by the CDF~II detector. Figure \ref{fig:zfb} shows the data compared with {\tt PYTHIA~v6.216} \cite{pythia}, indicating very good agreement across the entire $M_{ee}$ spectrum. The forward-backward asymmetry is sensitive to ${\rm sin}^2\theta_W^{eff}$, which is an effective parameter that includes higher order corrections. The current world average value of ${\rm sin}^2\theta_W^{eff}$ is 0.23145$\pm$0.00013 \cite{pdg}, the D0 collaboration measured ${\rm sin}^2\theta_W^{eff}=0.2326\pm0.0019$ from an earlier $Z$ forward-backward asymmetry measurement \cite{d0zfb} based on 1.1~fb$^{-1}$ of collision data and we plan to soon report ${\rm sin}^2\theta_W^{eff}$ from the CDF preliminary measurement presented here. 

\section{\boldmath $WW$, $WZ$, $ZZ$ diboson production}
The measurement of heavy diboson production in $p\bar{p}$ collisions is a way of probing the structure of the electroweak sector, as the presence of anomalous triple gauge bosons couplings for instance could indicate new physics at a higher mass scale. Moreover, as these processes are backgrounds to some Higgs boson decay signatures, their measurement provides milestones on the way to sensitivity to a Higgs boson production measurement. Heavy diboson production has been established at the Tevatron in several final states using 1--3~fb$^{-1}$ \cite{dibosond0,dibosoncdf}. As the collision data available increases, the precision of these measurements is being progressively improved. We report on a preliminary measurement of $WZ$ and $ZZ$ production in three- and four-lepton channels, using $\sim 6$~fb$^{-1}$ of data collected by the CDF experiment. From 50 $WZ\rightarrow\ell\nu$ candidate events and 4 $ZZ\rightarrow \ell\ell$ candidate events, we measure $\sigma(p\bar{p}\rightarrow WZ)=4.1\pm0.7~{\rm pb}$ and $\sigma(p\bar{p}\rightarrow ZZ)=1.7^{+1.2}_{-0.7}({\rm stat.})\pm0.2({\rm syst.})~{\rm pb}$. The systematic uncertainties are minimised by measuring the experimental ratio to the inclusive $\sigma(p\bar{p}\rightarrow Z)$ cross section and using the theoretical value of $\sigma(Z\rightarrow \ell\ell)$ production cross section. Figure \ref{fig:zzmass} shows the distribution of invariant mass for the $Z$ candidates reconstructed in the 4 $ZZ$ candidate events. The measurements are in agreement with the LO theoretical prediction of $\sigma(p\bar{p}\rightarrow WZ)=3.46\pm0.21$~pb and $\sigma(p\bar{p}\rightarrow ZZ)=1.4\pm0.1$~pb, and in agreement with earlier measurements at the Tevatron \cite{wzcdf,wzd0}. We also report on a preliminary measurement of combined $WW+WZ$ production cross section in the channel with an identified electron or muon, large missing transverse energy and two jets in 4.6~fb$^{-1}$ of CDF~II data. The analysis employs a matrix element technique which calculates event probability densities for signal and background hypotheses. The probabilities are used to form a discriminant variable which is evaluated for signal and background events. The resulting template distributions are fit to the data using a binned likelihood approach, and yield a cross section $\sigma_{WW+WZ}=16.5^{+3.3}_{-3.0}~{\rm pb}$. The  measurement increases the precision of previous determinations \cite{wwljetcdf,wwljetd0} and is in agreement with the theory prediction of 16.1$\pm$0.9~pb \cite{wwwztheo}.

\section{\boldmath Conclusions}
We have presented recent preliminary measurements at Tevatron of top pair and single production cross section, top quark mass and width, top pair spin correlations and forward-backward asymmetry. In the electroweak sector, we have presented a new Tevatron average of the $W$ boson width, and preliminary measurements of the $W$ and $Z$ forward-backward asymmetries and $WZ,ZZ$ diboson production cross sections. All measurements are based on larger amount of collision data than previously used and are in agreement with the SM.


\begin{thebibliography}{9}
\bibitem{wmt} The $W$ transverse mass is defined as $M_T=\sqrt{2(p_T^\ell p_T^\nu-\vec{p}^\ell_T\cdot\vec{p}_T^\nu)}$, where $\vec{p}_T^\ell$ and $\vec{p}_T^\nu$ are the measured transverse momentum of the charged lepton and the transverse momentum of the neutrino respectively.
\bibitem{rapidity} We assume ($z,~\phi,~\theta$) coordinate system with the $z$ axis in the direction of the proton beam; $\phi$ and $\theta$ are the azimuthal and polar angle, respectively. The rapidity ($y_W$) and the pseudorapidity ($\eta$) are defined as $y_W=ln\frac{E+P_z}{E-P_z}$ and $\eta=-{\rm ln}{\rm tan}\left(\frac{\theta}{2}\right)$.
\end{thebibliography}
\end{document}